\documentclass[11pt]{article}
\makeatletter\if@twocolumn\PassOptionsToPackage{switch}{lineno}\else\fi\makeatother
\pdfoutput=1
\usepackage{subcaption}
\usepackage{graphicx}

\usepackage[default]{jasa_harvard}
\let\citet\citeasnoun 
\let\citep\cite 

\usepackage{JASA_manu}

\usepackage{tabulary,amsmath,amssymb,amsfonts,amsbsy,xcolor}

\usepackage[T1]{fontenc}
\usepackage{lineno}
\usepackage{url,multirow,morefloats,floatflt,cancel,tfrupee}
\makeatletter

\AtBeginDocument{\@ifpackageloaded{textcomp}{}{\usepackage{textcomp}}}
\makeatother
\usepackage{colortbl}
\usepackage{xcolor}
\usepackage{pifont}
\usepackage[nointegrals]{wasysym}
\makeatletter

\def\mcWidth#1{\csname TY@F#1\endcsname+\tabcolsep}

\def\cAlignHack{\rightskip\@flushglue\leftskip\@flushglue\parindent\z@\parfillskip\z@skip}
\def\rAlignHack{\rightskip\z@skip\leftskip\@flushglue \parindent\z@\parfillskip\z@skip}

\@ifundefined{etal}{}{}

\usepackage{ifxetex}
\ifxetex\else\if@twocolumn\@ifpackageloaded{stfloats}{}{\usepackage{dblfloatfix}}\fi\fi

\AtBeginDocument{
\expandafter\ifx\csname eqalign\endcsname\relax
\def\eqalign#1{\null\vcenter{\def\\{\cr}\openup\jot\m@th
  \ialign{\strut$\displaystyle{##}$\hfil&$\displaystyle{{}##}$\hfil
      \crcr#1\crcr}}\,}
\fi
}

\AtBeginDocument{%
  \@ifpackageloaded{endfloat}%
   {\renewcommand\efloat@iwrite[1]{\immediate\expandafter\protected@write\csname efloat@post#1\endcsname{}}}{\newif\ifefloat@tables}%
}%

\def\BreakURLText#1{\@tfor\brk@tempa:=#1\do{\brk@tempa\hskip0pt}}
\let\lt=<
\let\gt=>
\def\processVert{\ifmmode|\else\textbar\fi}

\@ifundefined{subparagraph}{
\def\subparagraph{\@startsection{paragraph}{5}{2\parindent}{0ex plus 0.1ex minus 0.1ex}%
{0ex}{\normalfont\small\itshape}}%
}{}

\newcommand\role[1]{\unskip}
\newcommand\aucollab[1]{\unskip}
  
\@ifundefined{tsGraphicsScaleX}{\gdef\tsGraphicsScaleX{1}}{}
\@ifundefined{tsGraphicsScaleY}{\gdef\tsGraphicsScaleY{.9}}{}
\def\checkGraphicsWidth{\ifdim\Gin@nat@width>\linewidth
	\tsGraphicsScaleX\linewidth\else\Gin@nat@width\fi}

\def\checkGraphicsHeight{\ifdim\Gin@nat@height>.9\textheight
	\tsGraphicsScaleY\textheight\else\Gin@nat@height\fi}

\def\fixFloatSize#1{}%\@ifundefined{processdelayedfloats}{\setbox0=\hbox{\includegraphics{#1}}\ifnum\wd0<\columnwidth\relax\renewenvironment{figure*}{\begin{figure}}{\end{figure}}\fi}{}}
\let\ts@includegraphics\includegraphics

\def\inlinegraphic[#1]#2{{\edef\@tempa{#1}\edef\baseline@shift{\ifx\@tempa\@empty0\else#1\fi}\edef\tempZ{\the\numexpr(\numexpr(\baseline@shift*\f@size/100))}\protect\raisebox{\tempZ pt}{\ts@includegraphics{#2}}}}

\AtBeginDocument{\def\includegraphics{\@ifnextchar[{\ts@includegraphics}{\ts@includegraphics[width=\checkGraphicsWidth,height=\checkGraphicsHeight,keepaspectratio]}}}

\DeclareMathAlphabet{\mathpzc}{OT1}{pzc}{m}{it}

\def\URL#1#2{\@ifundefined{href}{#2}{\href{#1}{#2}}}

\def\UrlOrds{\do\*\do\-\do\~\do\'\do\"\do\-}%
\g@addto@macro{\UrlBreaks}{\UrlOrds}

\edef\fntEncoding{\f@encoding}

\makeatother

\newif\ifmultipleabstract\multipleabstractfalse%
\makeatletter\def\oupIndent{1pt}
\def\author#1{\gdef\@author{\hskip-\dimexpr(\tabcolsep)\hskip\oupIndent\parbox{\dimexpr\textwidth-\oupIndent}{\centering#1}}}

\makeatother

\usepackage[section]{placeins}

\usepackage{float}
\begin{document}

\title{Novelty and Primacy: A Long-Term Estimator for Online Experiments}

\author{Soheil Sadeghi, Somit Gupta, Stefan Gramatovici, Jiannan Lu, Hao Ai, Ruhan Zhang	\\	Microsoft}

\maketitle 

\newpage
\begin{center} \textbf{Abstract} \end{center}
Online experiments are the gold standard for evaluating impact on user experience and accelerating innovation in software. However, since experiments are typically limited in duration, observed treatment effects are not always permanently stable, sometimes revealing increasing or decreasing patterns over time. There are multiple causes for a treatment effect to change over time. In this paper, we focus on a particular cause, user-learning, which is primarily associated with novelty or primacy. Novelty describes the desire to use new technology that tends to diminish over time. Primacy describes the growing engagement with technology as a result of adoption of the innovation. User-learning estimation is critical because it holds experimentation responsible for trustworthiness, empowers organizations to make better decisions by providing a long-term view of expected impact, and prevents user dissatisfaction. In this paper, we propose an observational approach, based on difference-in-differences technique to estimate user-learning at scale. We use this approach to test and estimate user-learning in many experiments at Microsoft. We compare our approach with the existing experimental method to show its benefits in terms of ease of use and higher statistical power, and to discuss its limitation in presence of other forms of treatment interaction with time.  

\vspace*{.3in}\noindent\textsc{Keywords}: {A/B testing; difference-in-differences; user-learning; user experience; trustworthiness} %scale; statistical power
    
\newpage
\section{Introduction} \
\label{sec:introduction}
Online experiments (e.g., A/B tests) are the gold standard for evaluating impact on user experience in websites, mobile and desktop applications, services, and operating systems\unskip~\cite{kohavi2004emetrics,scott2010modern,tang2010overlapping,scott2015multi,urban2016s,kohavi2017online,kaufman2017democratizing,li2019experimentation,kohavi2020trustworthy}. Tech giants such as Amazon, Facebook, and Google invest in in-house experimentation systems, while multiple start-ups like Optimizely help other companies run A/B testing. At Microsoft, the experimentation system provides A/B testing solutions to many products including Bing, Cortana, Microsoft News, Office, Skype, Windows and Xbox, running thousands of experiments per year. The usefulness of controlled experiments comes from their ability to establish a causal relationship between the features being tested and the changes in user response. In the simplest controlled experiment or A/B test, users are randomly assigned to one of the two variants: control (A) or treatment (B). Usually control is the existing system, and treatment is the existing system with a new feature X. User interactions with the system are recorded and metrics are computed. If the experiment was designed and executed correctly, the only thing consistently different between the two variants is the feature X. External factors such as seasonality, impact of other feature launches, competitor moves, etc. are distributed randomly between control and treatment, and therefore do not impact the results of the experiment. Hence, any difference in metrics between the two groups must be due to the feature X. For online experiments where multiple features are tested simultaneously, more complicated designs are used to establish a causal relationship between the changes made to the product and changes in user response\unskip~\cite{haizler2020factorial,sadeghi2020sliced}. This is the key reason for widespread use of controlled experiments for evaluating impact on user experience for new features in software.

Having the right metrics is critical to successfully executing and evaluating an experiment\unskip~\cite{deng2016data,machmouchi2016principles}. The overall evaluation criteria metric plays a key role in the experiment to make a ship/no-ship decision\unskip~\cite{kohavi2009controlled}. The metric changes observed during the experiment (typically few weeks or few months) are not always permanently stable, sometimes revealing increasing or decreasing patterns over time. There are multiple causes for a treatment effect to change over time. In this paper, we focus on one particular cause, user-learning, which was first proposed in Thorndike's Law of Effect\unskip~\cite{thorndike1898animal}. According to this law, positive and negative outcomes reinforce the behaviors that caused them. User-learning and statistical modeling first came together in the 50s\unskip~\cite{estes1950toward,bush1951mathematical}. In online experiments with changing treatment effect over time, user-learning is primarily associated with novelty or primacy effect. Novelty effect describes the desire to use new technology that tends to diminish over time. On the contrary, primacy effect describes the growing engagement with technology as a result of adoption of the innovation. These effects have been discussed in multiple fields by many studies\unskip~\cite{anderson1961primacy,peterson1967primacy,jones1968pattern,bartolomeo1997novelty,tan2000recency,howard2003empirical,feddersen2006novelty,poppenk2010revisiting,li2010primacy,kohavi2012trustworthy,mutsuddi2012text,hohnhold2015focusing,van2016first,dmitriev2016pitfalls,belton2018attention,chen2019b}.

User-learning estimation and understanding the sustained impact of the treatment effect is critical for many reasons. First, it holds experimentation responsible for preventing overestimation or underestimation in the case of novelty or primacy. Second, it empowers organizations to make better decisions by providing them a long-term view of expected changes in the key metrics. Often, experiments show gain in one key product metric and loss in another. In this case, the product owners need to trade-off two metrics to make a ship decision. This can lead to a wrong decision if the sustained treatment effect is different from the observed treatment effect.  Third, it ensures that the experiment is not causing user dissatisfaction even though the key metrics might have moved in the positive direction. At times, undesirable treatments, that cause distraction or confusion among users, may initially lead to an increase in some metrics indicative of higher engagement. For instance, in an experiment shared in\unskip~\citet{dmitriev2017dirty}, there was a bug in treatment which led to users seeing a blank page. This resulted in a huge spike in the number of impressions from that page, as users tried to refresh the page multiple times to see if that would help them see any contents on the page.

To motivate this paper, let us consider an experiment from\unskip~\citet{dmitriev2017dirty} on the Microsoft News homepage where the treatment replaced the Outlook.com button with the Mail app button on the top stripe (msn.com experiment in Figure~\ref{f-5dabc0fe04f5}). The experiment showed a 4.7\% increase in overall clicks on the page, a 28\% increase in the number of clicks on the button, and a 27\% increase in the number of clicks on the button adjacent to the Mail app button. Ignoring any concerns about novelty effect, this would seem like a great result.

\bgroup
\fixFloatSize{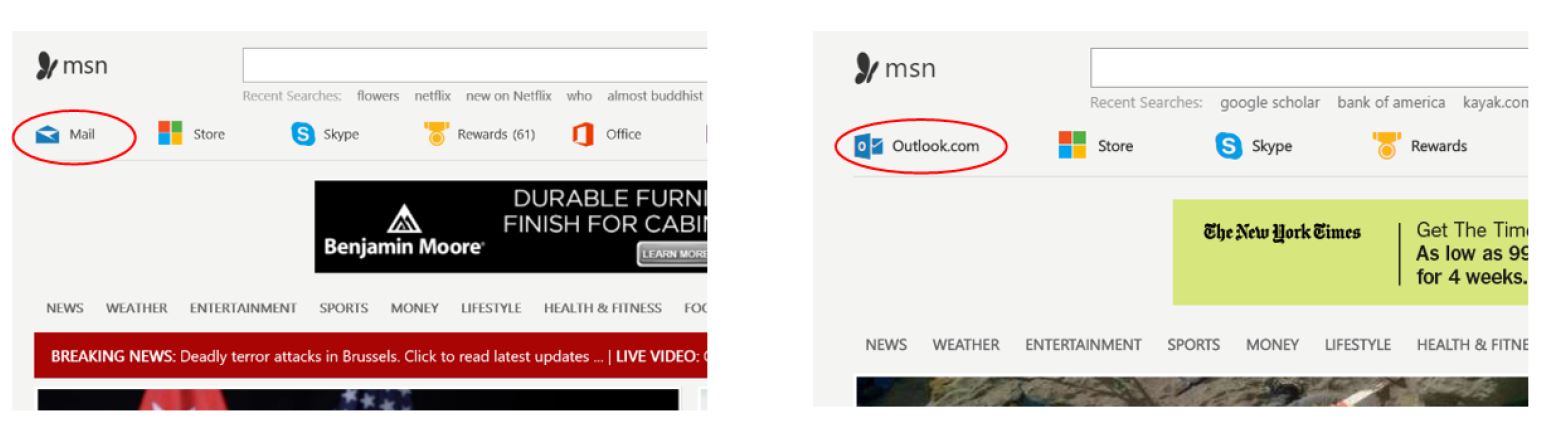}
\begin{figure}[!htbp]
\centering \makeatletter\IfFileExists{Figures/outlookexperiment.jpg}{\includegraphics{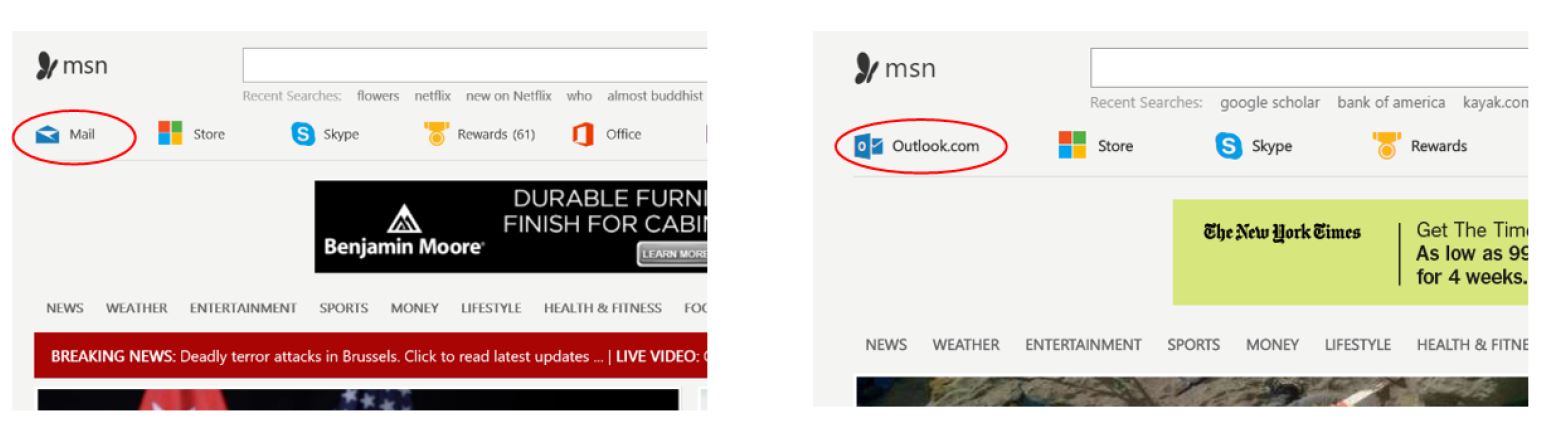}}{}
\makeatother 
\caption{{Screenshots of treatment with the Mail app button (left), and control with the Outlook.com button (right) in the msn.com experiment.}}
\label{f-5dabc0fe04f5}
\end{figure}
\egroup

However, novelty effect likely exists in this experiment. Looking at each day segment, we found that the difference between number of clicks on the Mail app (in treatment) and Outlook.com (in control) were decreasing rapidly day over day (see Figure~\ref{f-3f004c6957a2}). We believe that the treatment caused a lot of confusion to the users who were used to navigating to Outlook.com from the msn.com page. When the button instead started opening the Mail app, some users continued to click on the button expecting it to work like it used to. They may have also clicked on the button adjacent to Mail app button to check if other buttons work. Overtime, users learned that the button has changed and stopped clicking on it. Had this treatment been shipped to all users, it could have caused a lot of user dissatisfaction. In fact, we shut down the experiment mid-way to avoid user dissatisfaction.

\bgroup
\fixFloatSize{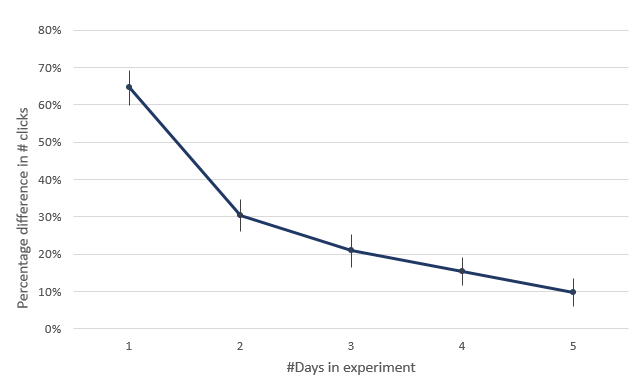}
\begin{figure}[!htbp]
\centering \makeatletter\IfFileExists{Figures/outlookexperimentdaybyday.png}{\includegraphics[width=.6\linewidth]{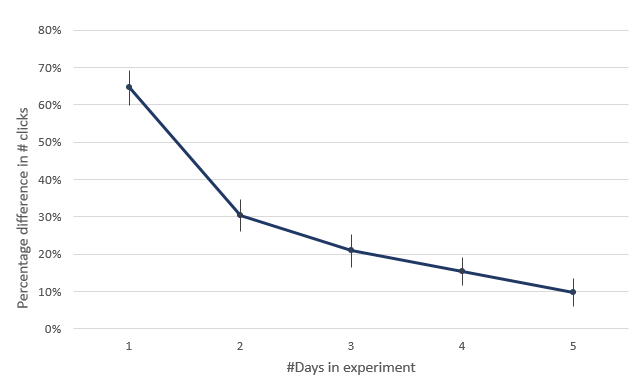}}{}
\makeatother 
\caption{{The percentage difference in number of clicks on the Mail/Outlook.com button, each day, between treatment and control in the msn.com experiment.}}
\label{f-3f004c6957a2}
\end{figure}
\egroup

In this case, we had a sound hypothesis for the cause of novelty effect. This hypothesis fits well with many observations made above. However, we may not be that fortunate for the hundreds of other experiments, or for many other key business metrics. At Microsoft, we run experimentation at a large scale. Typically, more than a 1000 online experiments with tens (sometimes hundreds) of metrics are run at Microsoft each month\unskip~\cite{gupta2019b}. Therefore, we need methods for estimating, testing, and utilizing user-learning to estimate the long-term impact of feature changes at scale. 

The methodology currently used in industry for user-learning estimation is based on an experimental approach first proposed by\unskip~\citet{hohnhold2015focusing}. This approach provides an unbiased estimate of user-learning by adding a significant operational changes to the experimentation system. It also requires a large pool of experimental units at the beginning of the experiment to be randomly divided into multiple cohorts and to be assigned to treatment in a ladder form. This approach is usually used for select few experiments where the feature team suspects user-learning a priori and is willing to have a complex experimental design setting to estimate it. It is practically more effective to estimate user-learning without any changes in the experimentation system.

In this paper, we propose an observational approach, based on the well-known difference-in-differences technique\unskip~\cite{abadie2005semiparametric,athey2006identification,donald2007inference,conley2011inference,dimick2014methods} to estimate user-learning at scale. We use this approach to detect user-learning in many experiments at Microsoft. Our formulation is powerful in quickly testing for the presence of user-learning even in short duration experiments. The main advantage of our proposed methodology is that it provides a practically more effective way to estimate user-learning by eliminating the need for the experimental design setting required by\unskip~\citet{hohnhold2015focusing}. Additionally, our proposed approach provides more statistical power for testing the significance of user-learning compared to the existing approach. We further illustrate this with a simulation study. The main disadvantage of our proposed methodology is that, although it provides an unbiased estimate of the long-term treatment effect, user-learning estimation is more susceptible to other forms of treatment interaction with time (e.g., seasonality). Practically in controlled experiments, having a large treatment and seasonality interaction effect that significantly biases user-learning estimation is rare. Further, more advanced techniques such as time series decomposition of seasonality can be used to reduce the bias in user-learning estimation.

In general, we recommend using observational approach to test for the presence of user-learning. In the case where user-learning is significant, we usually recommend running the experiment longer to allow for the treatment effect to stabilize\unskip~\cite{dmitriev2016pitfalls}. If the user-learning is gradually changing over time, we recommend running the experiment long enough and utilizing observational approach to construct user-learning time series that can be extrapolated to estimate the long-term treatment effect. In cases, where we suspect strong seasonality interaction with the treatment effect, and the feature team is willing to use a larger sample size with more complicated setting, experimental approach can be useful.

The remainder of the paper is organized as follows. In Section~\ref{sec:formulation}, we first formulate the problem and discuss a natural way to visually check for the presence of user-learning. In Section~\ref{sec:existingMethodology}, we review the existing experimental approach for user-learning estimation. Next, we propose a new observational approach, based on difference-in-differences technique, and compare our methodology to the existing method in Section~\ref{sec:proposedMethodology}. In Section~\ref{sec:application}, we illustrate user-learning estimation using another Microsoft experiment and a simulation study. We conclude in Section~\ref{sec:conclusion}.

%\clearpage
\section{Formulation} \
\label{sec:formulation}
Without loss of generality, let us consider an A/B test in which $n$ experimental units (e.g., browser cookies, devices, etc.) are randomly divided in two cohorts based on the hash of experimental unit id\unskip~\cite{kohavi2017online}. We assign one cohort to control and the other to treatment. Let the experiment duration consists of $k-1$ time windows (e.g., days, weeks, months, etc.). We usually have one or multiple weeks to account for day-of-the-week effects\unskip~\cite{kohavi2020trustworthy,dmitriev2016pitfalls}. For a metric of interest $y$, we define $C^{j}$ and $T^{j}$ to be the sample mean $\overline{y}$ for the control and treatment cohorts in time window $j$,  $j=1, \cdots, k-1$ , respectively. For a given $t$ , where $t=1, \cdots, k-1$ , we define the estimated treatment effect, ${\widehat{\tau}}_t$, as follows:
\begin{equation}
\label{eq-treatmenttimeseries}
{\widehat{\tau}}_t=\sum_{j\leq t}\frac1t(T^{j}-C^{j}).
\end{equation}
Figure~\ref{f-ac77dc49c80f} visually displays the aforementioned A/B test.

\bgroup
\fixFloatSize{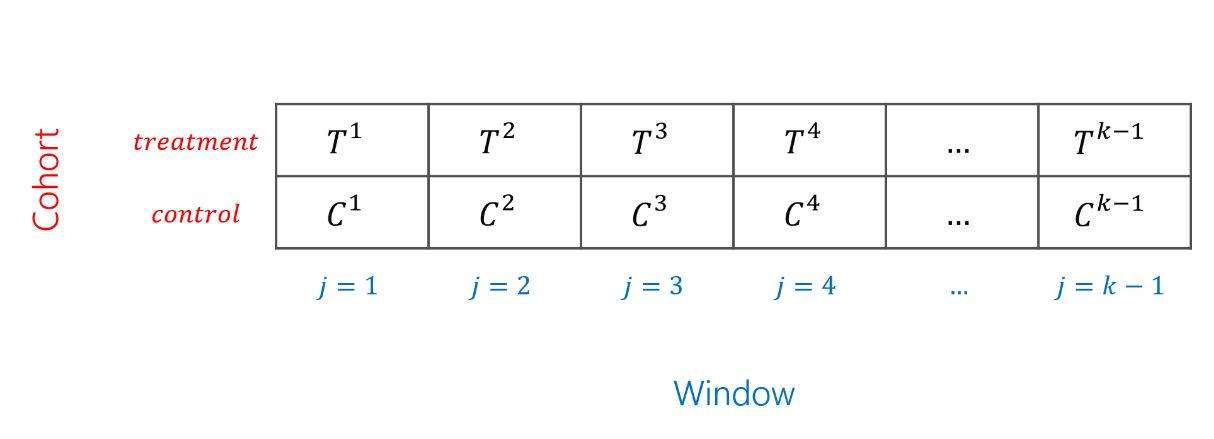}
\begin{figure}[!htbp]
\centering \makeatletter\IfFileExists{Figures/abtesting.jpg}{\includegraphics[width=.6\linewidth]{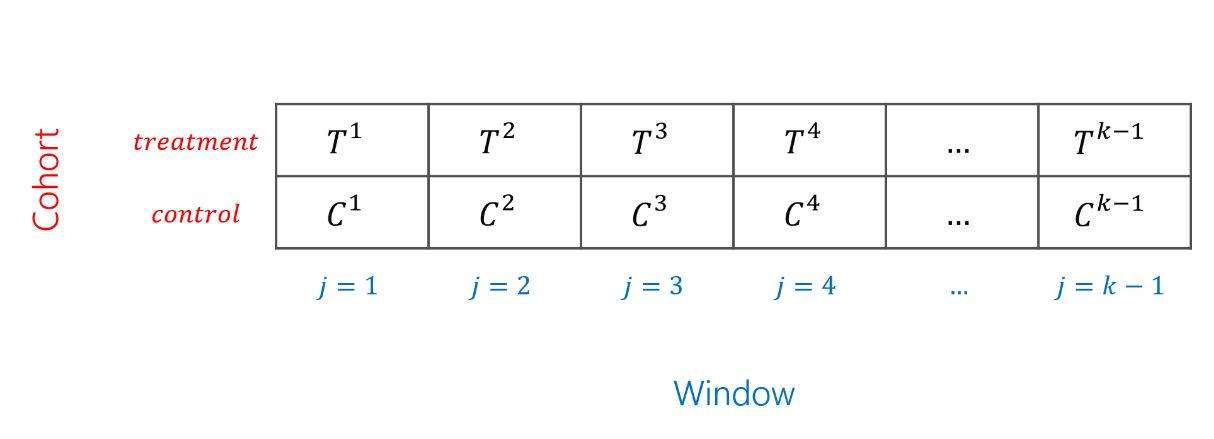}}{}
\makeatother 
\caption{{A/B test during $k-1$ time windows}}
\label{f-ac77dc49c80f}
\end{figure}
\egroup

Note that different experimental units may get exposed to the experiment at different time windows. Further, after being exposed, there may be time windows where these units do not use the product. For the experimental units that are missing for some time windows, we can take two approaches in treatment effect estimation. The first approach is to impute zero for missing values in metrics where missing can be interpreted as zero, e.g., number of clicks, time spent, and dollar amount. This approach allows us to use all experimental units in all time windows for the analysis which benefits from the high statistical power in testing effect significance. However, for situations where imputation is not feasible, in each time window, we only include the experimental units for which we observe the metric. In this case, we assume that the missing value distributions are consistent between the control and treatment cohorts. Practically, this assumption is feasible for majority of randomized experiments\footnote{This depends on the metric being computed. Metrics that do not include all users are more likely to be affected by sample ratio mismatch. There may also be some cases where the treatment impacts the propensity of a unit to return to a product more (or less) often leading to a sample ratio mismatch in a time window \unskip~\cite{fabijan2019diagnosing}.}.

The A/B test in Figure~\ref{f-ac77dc49c80f} is limited in duration (typically few weeks or few months), and the observed treatment effect $\widehat{\tau}_t$ is not always permanently stable, sometimes revealing increasing or decreasing patterns over time. There are multiple causes for a treatment effect to change over time. In this paper, we focus on one particular cause, user-learning, which was first proposed in Thorndike's Law of Effect\unskip~\cite{thorndike1898animal}. According to this law, positive and negative outcomes reinforce the behaviors that caused them. User-learning and statistical modeling first came together in the 50s\unskip~\cite{estes1950toward,bush1951mathematical}. In online experiments with changing treatment effect over time, user-learning is primarily associated with novelty or primacy effect. Novelty effect describes the desire to use new technology that tends to diminish over time (see Figure~\ref{f-ecf0d8992112}). On the contrary, primacy effect describes the growing engagement with technology as a result of adoption of the innovation (see Figure~\ref{f-fed3d515e40c}). These effects have been discussed by multiple studies in the online experimentation literature\unskip~\cite{kohavi2012trustworthy,hohnhold2015focusing,dmitriev2016pitfalls,chen2019b}.

\bgroup
\begin{figure}[!htbp]
 \centering
  \centering
 \begin{subfigure}[t]{0.49\linewidth}
   \includegraphics[width=.8\linewidth]{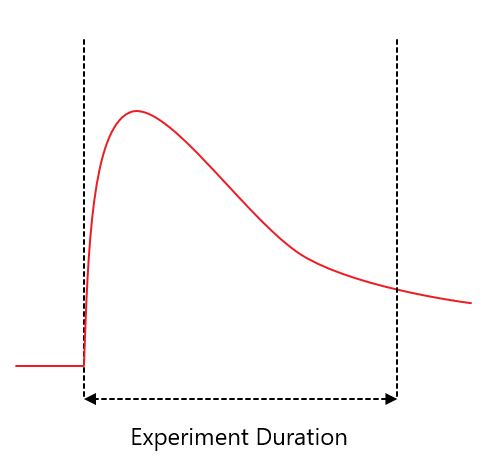}
   \caption{Novelty effect}
   \label{f-ecf0d8992112}
 \end{subfigure}
 \begin{subfigure}[t]{0.49\linewidth}
   \includegraphics[width=.8\linewidth]{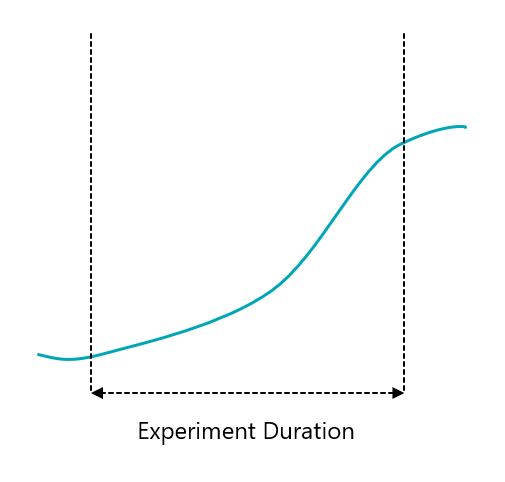}
   \caption{Primacy effect}
   \label{f-fed3d515e40c}
 \end{subfigure}
 \caption{A/B test with novelty or primacy effect}
\end{figure}
\egroup

For an experimental unit that gets exposed to treatment at time window $t_0+1$, define $\delta_{t-t_0}$ to be the user learned effect from the $(t_0+1)^{\textrm{st}}$ time window to the $t^{\textrm{th}}$ time window. By this definition $\delta_t$ is the user learned effect from the first time window ($t_0 = 0$) to the $t^{\textrm{th}}$ time window and $\delta_1 = 0$. Let us assume that $\mathrm{E}(y_t) = \mu_t$ has a linear form,
\begin{equation}
\label{eq-userlearningassumption}
{\mu}_t = \alpha + \beta_t + (\tau + \delta_{t-t_0})I_\tau,
\end{equation}
where $\alpha$ is the intercept, $\beta_t$ is the $t^{\textrm{th}}$ time window main effect, $\tau$ is treatment main effect, $\delta_{t-t_0}$ is user-learning, and $I_\tau$ is an indicator which equals to 1 if the metric is measured in the treatment cohort. This is a reasonable assumption as in cases where there are nonlinear effects, this model can be considered as a first order approximation of the Taylor series expansion.

Detecting user-learning and understanding the long-term treatment effect is critical while making a ship decision. The most intuitive approach is to look at the time series $\widehat{\tau}_1, \widehat{\tau}_2, \cdots, \widehat{\tau}_{k-1}$ and see if there exists an increasing or decreasing pattern\unskip~\cite{chen2019b}. If the time series treatment effect is permanently stable (see Figure~\ref{f-d993292f9212}), any $\widehat{\tau}_t$, where $t=1, \cdots, k-1$, can be viewed as the long-term impact of the feature change. If there exist an increasing or decreasing pattern and the pattern has converged at time $t < k-1$ (see Figure~\ref{f-d67d446f6281}), then $\widehat{\tau}_{k-1}$ can be viewed as the long-term impact of the feature change. However, for the situations where there exists an increasing or decreasing pattern and it has not converged during the experiment, we cannot quantify the long-term effect simply by looking at the time series $\widehat{\tau}_1, \widehat{\tau}_2, \cdots, \widehat{\tau}_{k-1} $. To illustrate this, we revisit the msn.com experiment in Figure~\ref{f-5dabc0fe04f5}.

\bgroup
\begin{figure}[!htbp]
 \centering
  \centering
 \begin{subfigure}[t]{0.49\linewidth}
   \includegraphics[width=.8\linewidth]{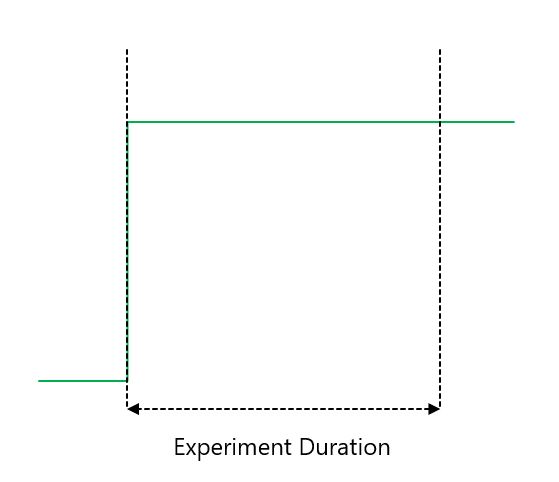}
   \caption{Permanently stable}
   \label{f-d993292f9212}
 \end{subfigure}
 \begin{subfigure}[t]{0.49\linewidth}
   \includegraphics[width=.8\linewidth]{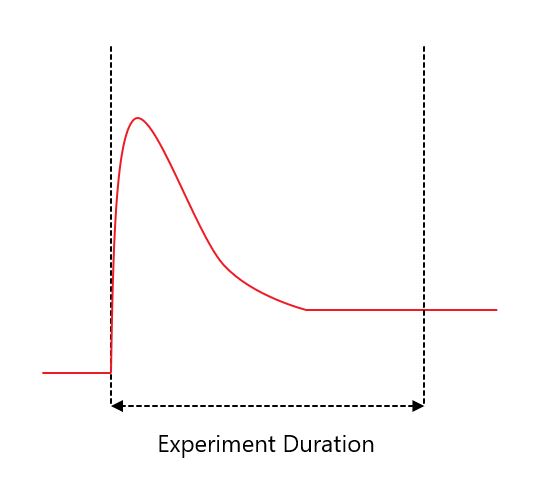}
   \caption{Decreasing pattern with convergence}
   \label{f-d67d446f6281}
 \end{subfigure}
 \caption{Behavior of treatment effect estimate during the A/B test}
\end{figure}
\egroup

Figure~\ref{f-d982ca7a0d8e} shows the time series $\widehat{\tau}_1, \widehat{\tau}_2, \cdots, \widehat{\tau}_{k-1}$ and the corresponding confidence interval over each day of the experiment for total number of clicks. In this Figure, the confidence interval on the first day overlaps with the confidence interval on almost every other day. Further, the treatment effect is significant on first day (0.244, [0.129, 0.360]), tends to decline and become insignificant in the next few days, and becomes significant again on the last day. Therefore, there is a need for a more rigorous statistical approach to test the significance of user-learning.

\bgroup
\begin{figure}[!htbp]
 \centering
  \centering
 \begin{subfigure}[t]{0.49\linewidth}
   \includegraphics[width=.9\linewidth]{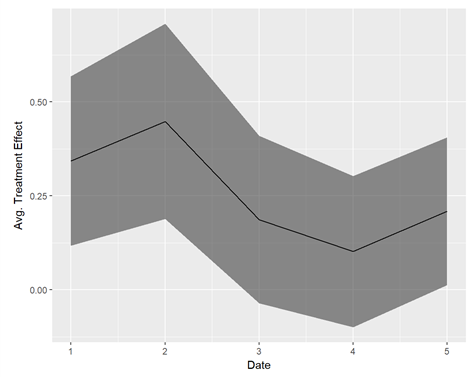}
   \caption{Calendar date}
   \label{f-d982ca7a0d8e}
 \end{subfigure}
 \begin{subfigure}[t]{0.49\linewidth}
   \includegraphics[width=.9\linewidth]{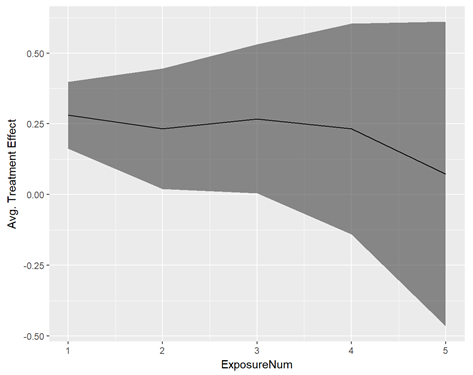}
   \caption{Exposure days}
   \label{f-de0f86e7f90d}
 \end{subfigure}
 \caption{Estimated treatment effect time series for total number of clicks in the msn.com experiment.}
 \label{f-totalclicks}
\end{figure}
\egroup

One of the factors in treatment effect fluctuation of Figure~\ref{f-d982ca7a0d8e} is that users are exposed to the experiment at different dates. Thus, it may seem more appropriate to provide visualization based on days of exposure. Figure~\ref{f-de0f86e7f90d} shows the estimated treatment effect for total number of clicks based on exposure days. Although this graph better visually conveys the presence of novelty, it does not yet provide a statistical test for its significance. First, the confidence interval on the first exposure day overlaps with the confidence interval on almost all other exposure days. Second, number of users with $t$ exposure days decreases as $t$ increases. This leads to increase in confidence intervals for higher exposure days which makes it harder to detect novelty. Third, the set of users who have 5 exposure days are likely the most engaged/loyal users. The presence of any heterogeneous treatment effect that interacts with the loyalty level of users can lead to increasing or decreasing trend which is not related to user-learning\unskip~\cite{wang2019heavy}. Studying the sub-population estimates of the treatment effect comes with its own pitfalls: it is not representative of the entire user base\unskip~\cite{wang2019heavy}, and it will have lower statistical power to detect a change due to decrease in sample size\unskip~\cite{kohavi2009controlled}. Therefore, we need a methodology to estimate user-learning and to statistically test for its significance. In the next section, we review an existing experimental approach used in industry to tackle user-learning.

%\clearpage
\section{Existing Methodology} \
\label{sec:existingMethodology}
In this section, we review an experimental design approach that was proposed by\unskip~\citet{hohnhold2015focusing} for user-learning estimation of the A/B test in Figure~\ref{f-ac77dc49c80f} (remember that A/B test duration consists of $k-1$ time windows). The purpose of this approach is to create a time series that provides an unbiased estimate of $\delta_t$. For simplicity, we first develop the concept in the case where the experiment duration is divided into two time windows ($k=3$) and show why the estimated user-learning is unbiased. We then expand the concept to the more general case with $k > 3$.

\subsection{Experimental Approach to User-Learning with $k=3$} \
For developing the concept in the simplest setup where the experiment duration consists of two time windows ($k=3$), we randomly divide the $n$ experimental units into $3$ cohorts. We assign the first cohort to control, and the second cohort to treatment. The third cohort's assignment switches from control to treatment in the second time window. We use index $i$ to refer to cohort, index $j$ to refer to time window, and $T_i^{j}$ or $C_i^{j}$ to refer to sample mean $\overline{y}$ of the cohort $i$ in time window $j$ if it is assigned to treatment or control, respectively (see Figure~\ref{f-5ea855b5b392}). Following Equation~(\ref{eq-treatmenttimeseries}), we can estimate the treatment effect time series with $\widehat{\tau}_1 = T_2^{1} - C_1^{1}$ and $\widehat{\tau}_2 = \frac{1}{2}[ (T_2^{1} - C_1^{1}) + (T_2^{2} - C_1^{2}) ]$. To estimate $\delta_2$, define $\widehat{\delta}_2$ as follows:
\begin{equation}
\label{eq-delta2experimental}
\widehat{\delta}_2 = T_2^{2} - T_3^{2}.
\end{equation}

\bgroup
\fixFloatSize{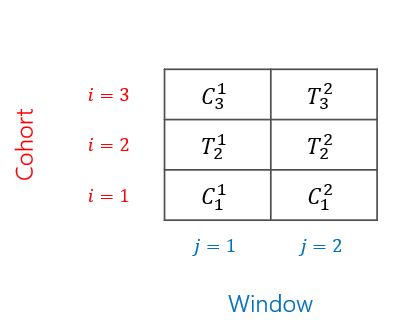}
\begin{figure}[!htbp]
\centering \makeatletter\IfFileExists{Figures/experimental_approach_k3.jpg}{\includegraphics[width=.4\linewidth]{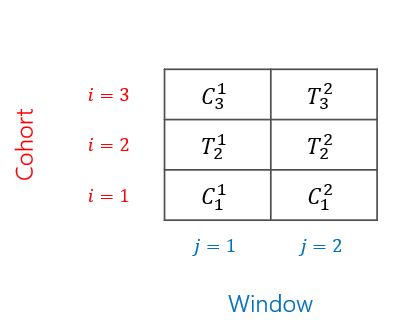}}{}
\makeatother 
\caption{{A/B test with three cohorts and two time windows}}
\label{f-5ea855b5b392}
\end{figure}
\egroup
 
Next, we show that $\widehat{\delta}_2$ in Equation~(\ref{eq-delta2experimental}) is an unbiased estimate of $\delta_2$. Remember that Equation~(\ref{eq-userlearningassumption}) poses a linear form on $\mu_t$ where ${\mu}_t = \alpha + \beta_t + (\tau + \delta_{t-t_0})I_\tau$. Since the second cohort is assigned to treatment on the first time window,
\begin{equation}
\mathrm{E}(T_2^{2}) = \alpha + \beta_2 + \tau + \delta_2.
\end{equation}
In addition, since the third cohort is assigned to treatment on the second time window,
\begin{equation}
\mathrm{E}(T_3^{2}) = \alpha + \beta_2 + \tau.
\end{equation}
Therefore,
\begin{equation}
\label{eq-delta2expectationexperimental}
\mathrm{E}(\widehat{\delta}_2) = \mathrm{E}(T_2^{2} - T_3^{2}) = \delta_2.
\end{equation}
In other words, $T_3^{2}$ and  $T_2^{2}$ are similar in all respects except for the fact that experimental units in $T_2^{2}$ have been exposed to treatment for a longer time period compared to those of $T_3^{2}$. If there are statistically significant differences between $T_2^{3}$ and $T_2^{2}$, then we can attribute the difference to user-learning.

\subsection{Experimental Approach to User-Learning with $k>3$}\
To expand the concept to the more general case with $k > 3$, we randomly divide the $n$ experimental units into $k$ cohorts (the intent is to have as many cohorts to cover all $k-1$ time windows). We assign the first cohort to control, the second cohort to treatment, and denote $T_i^{j}$ or $C_i^{j}$ to refer to sample mean $\overline{y}$ of the cohort $i$ in time window $j$ if it is assigned to treatment or control, respectively. Following Equation~(\ref{eq-treatmenttimeseries}), we can estimate the treatment effect time series with ${\widehat{\tau}}_t=\sum_{j\leq t}\frac1t(T_2^{j}-C_1^{j}) $, where $t=1, \cdots, k-1$. To estimate the user-learning, we switch the assignment of cohort $i\geq3$ from control to treatment in a ladder form (see Figure~\ref{f-23ecc4bedcd4}). We then estimate $\delta_t$ as follows:
\begin{equation}
\label{eq-deltaexperimental}
\widehat{\delta}_{t} = T_2^{t} - T_{t+1}^{t}.
\end{equation}

\bgroup
\fixFloatSize{Figures/experimental_approach_k.png}
\begin{figure}[!htbp]
\centering \makeatletter\IfFileExists{Figures/experimental_approach_k.png}{\includegraphics[width=.6\linewidth]{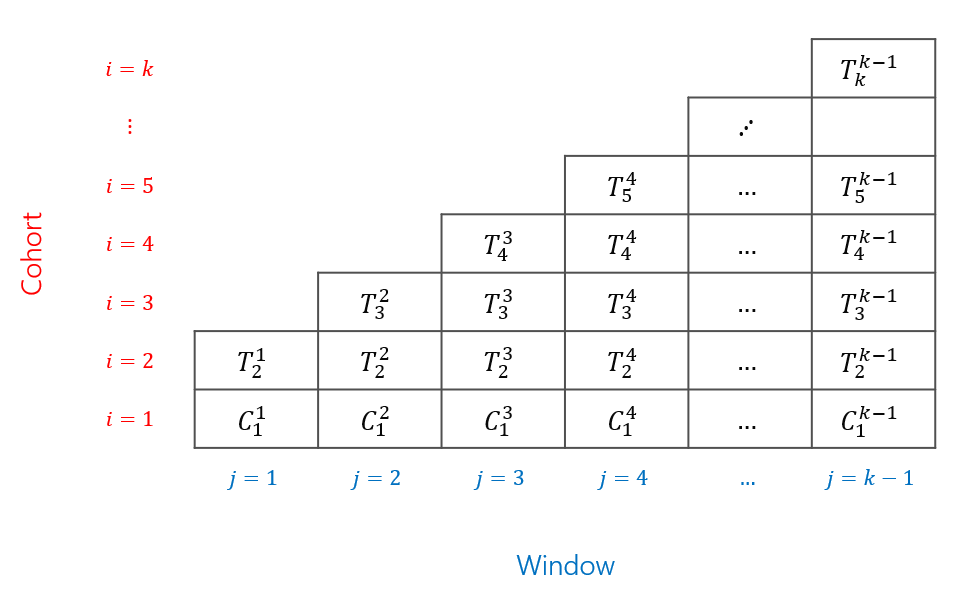}}{}
\makeatother 
\caption{{A/B test with $k$ cohorts and $k-1$ time windows}}
\label{f-23ecc4bedcd4}
\end{figure}
\egroup

Using Equation~(\ref{eq-userlearningassumption}), a similar argument can show that $\widehat{\delta}_{t}$ in Equation~(\ref{eq-deltaexperimental}) is an unbiased estimate of $\delta_{t}$. However, this estimator is not unique and other calculations can also provide an unbiased estimate of $\delta_{t}$. For example, we can use a cross-sectional approach in the same setup to construct the estimated user-learning time series. In this approach, $\delta_t$ is estimated with,
\begin{equation}
\label{eq-deltaexperimentalalt}
\widehat{\delta}_{t} = T_{k-t+1}^{k-1} - T_k^{k-1}.
\end{equation}

In addition to the time series $\widehat{\tau}_1, \widehat{\tau}_2, \cdots, \widehat{\tau}_{k-1}$ for treatment effect, this process also provides us a time series $\widehat{\delta}_2, \widehat{\delta}_3, \cdots, \widehat{\delta}_{k-1}$ to estimate user-learning.
However, it adds a significant operational changes to the experimentation system. It also requires a large pool of experimental units at the beginning of the experiment to be randomly divided into multiple cohorts and to be assigned to treatment in a ladder form. This approach is usually used for select few experiments where the feature team suspects user-learning a priori and is willing to have a complex experimental design setting to estimate it. It is practically more effective to estimate user-learning without any changes in the experimentation system. In the next section, we propose an observational approach, based on difference-in-differences to estimate user-learning at scale. The purpose of our approach is to eliminate the need for the aforementioned experimental design setting. we use this approach to estimate user-learning for experiments at Microsoft.

%\clearpage
\section{Proposed Methodology} \
\label{sec:proposedMethodology}
In this section, we propose an observational approach for user-learning estimation of the A/B test in Figure~\ref{f-ac77dc49c80f} (remember that A/B test duration consists of $k-1$ time windows). Our proposed approach is based on the well-known difference-in-differences (DID) technique\unskip~\cite{abadie2005semiparametric,athey2006identification,donald2007inference,conley2011inference,dimick2014methods}. We use DID to detect user-learning in many experiments at Microsoft. Our formulation is powerful in quickly testing for the presence of user-learning even in short duration experiments. Similar to the prior section, we first develop the concept in the case where the experiment duration is divided into two time windows ($k=3$) and show why the estimated user-learning is unbiased. We then expand the concept to the more general case with $k > 3$. The purpose of this approach is to estimate user-learning without the need for the experimental design setting we discussed in the prior section. Later in this section, we show that our proposed approach provides more statistical power for testing the significance of user-learning compared to the existing approach. We also discuss its drawback, and provide guidance on how to mitigate it practically.

\subsection{Observational Approach to User-Learning with $k=3$} \
For developing the concept in the simplest setup where the experiment duration consists of two time windows ($k=3$), we randomly divide the $n$ experimental units into $2$ cohorts. We assign the first cohort to control, and the second cohort to treatment. Following Equation~(\ref{eq-treatmenttimeseries}), we can estimate the treatment effect time series with $\widehat{\tau}_1 = T^{1} - C^{1}$ and $\widehat{\tau}_2 = \frac{1}{2}[ (T^{1} - C^{1}) + (T^{2} - C^{2}) ]$. To estimate $\delta_2$, define $\widehat{\delta}_2$ as follows:
\begin{equation}
\label{eq-delta2observational}
\widehat{\delta}_2 = (T^{2}-T^{1}) - (C^{2}-C^{1}).
\end{equation}

\bgroup
\fixFloatSize{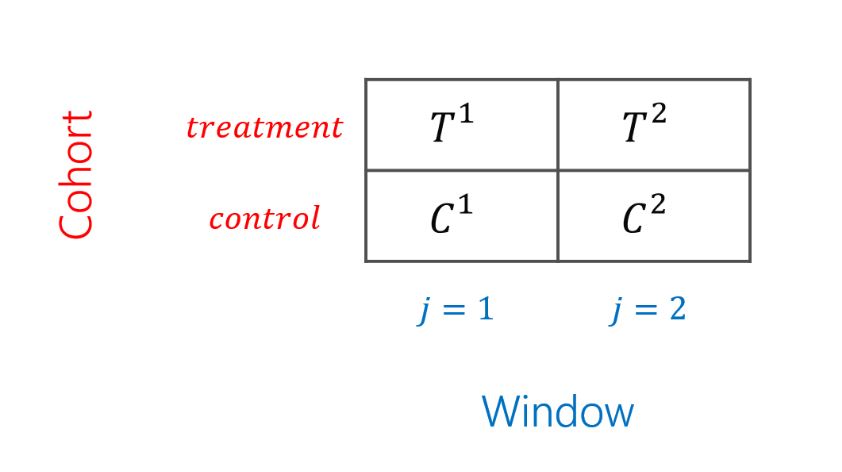}
\begin{figure}[!htbp]
\centering \makeatletter\IfFileExists{Figures/observational_approach_k3.jpg}{\includegraphics[width=.4\linewidth]{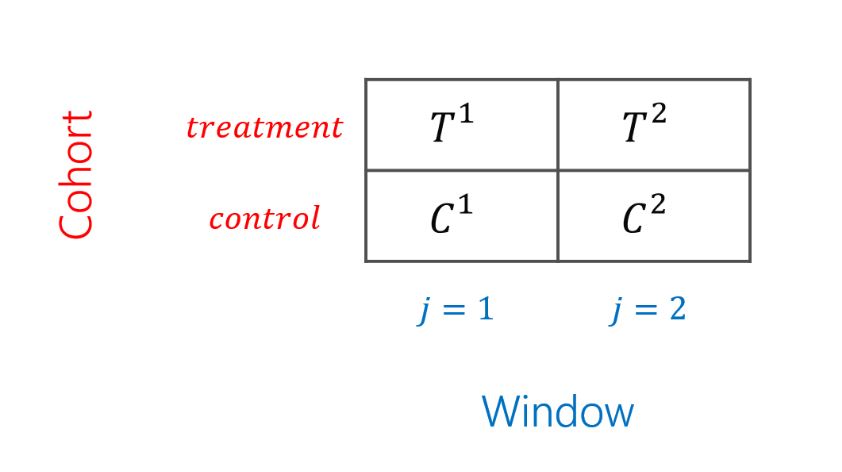}}{}
\makeatother 
\caption{{A/B test with two cohorts and two time windows}}
\label{f-577bb797b394}
\end{figure}
\egroup

Next, we show that $\widehat{\delta}_2$ in Equation~(\ref{eq-delta2observational}) is an unbiased estimate of $\delta_2$. Remember that Equation~(\ref{eq-userlearningassumption}) poses a linear form on $\mu_t$ where ${\mu}_t = \alpha + \beta_t + (\tau + \delta_{t-t_0})I_\tau$. Since the second cohort is assigned to treatment on the first time window,
\begin{equation}
\begin{aligned}
\mathrm{E}(T^{2}) &= \alpha + \beta_2 + \tau + \delta_2 \\
\mathrm{E}(T^{1}) &= \alpha + \beta_1 + \tau.
\end{aligned}
\end{equation}
Further, since the first cohort is assigned to control on the first time window,
\begin{equation}
\begin{aligned}
\mathrm{E}(C^{2}) &= \alpha + \beta_2 \\
\mathrm{E}(C^{1}) &= \alpha + \beta_1.
\end{aligned}
\end{equation}
Therefore,
\begin{equation}
\label{eq-deltaexpectationobservational}
\mathrm{E}(\widehat{\delta}_2) = \mathrm{E}(T^{2}-T^{1}) - \mathrm{E}(C^{2}-C^{1}) = \delta_2.
\end{equation}

\subsection{Observational Approach to User-Learning with $k>3$} \
Here we use the exact setup of the A/B test in Figure~\ref{f-ac77dc49c80f}, and estimate $\delta_t$ with,
\begin{equation}
\label{eq-deltaobservational}
\widehat{\delta}_t = (T^{t}-T^{1}) - (C^{t}-C^{1}).
\end{equation}
Using Equation~(\ref{eq-userlearningassumption}), a similar argument can show that $\widehat{\delta}_{t}$ in Equation~(\ref{eq-deltaobservational}) is an unbiased estimate of $\delta_{t}$. In addition to the time series $\widehat{\tau}_1, \widehat{\tau}_2, \cdots, \widehat{\tau}_{k-1}$ for the treatment effect, this approach also provides us a time series $\widehat{\delta}_2, \widehat{\delta}_3, \cdots, \widehat{\delta}_{k-1}$ to estimate user-learning. In fact, this approach provides estimation without any changes in the experimentation system.

\subsection{Statistical Test and Long-Term Estimation} \
The advantage of creating a time series to estimate user-learning is that we can statistically test its significance in the face of new technology, and utilize it to estimate the long-term impact of the treatment effect. To statistically test for the significance of user-learning, under Gaussian assumption, we can construct the $(1-\alpha)$ confidence interval $\widehat{\delta}_t \pm \phi^{-1}(1-\alpha/2) * \widehat{\mathrm{Var}}(\widehat{\delta}_t)$ where $\phi^{-1}$ is the normal inverse cumulative distribution function. User-learning is statistically significant if zero is outside the confidence interval. Next, we compare $\mathrm{Var}(\widehat{\delta}_t)$ from the experimental and observational approaches to see which approach provides more statistical power. For the metric of interest $y$, let us assume that the variance of each experimental unit within each time window is $\sigma ^{2}$. Let us also assume that the correlation of the metric for each experimental unit in two time windows is $\rho > 0$. To include all the experimental units in the analysis, we impute the missing values with zero (assuming the metric of interest allows for such interpretation of missing values). For simplicity, let us also assume that $n$ experimental units are equally divided between cohorts in both experimental and observational approaches. For the experimental approach, since each cohort includes $n/k$ experimental units,
\begin{equation}
\begin{aligned}
\mathrm{Var}(T_{2}^{t}) &= k\sigma ^{2} / n \\
\mathrm{Var}(T_{t+1}^{t}) &= k\sigma ^{2} / n.
\end{aligned}
\end{equation}
Thus, the variance of $\widehat{\delta}_t$ in Equation~(\ref{eq-deltaexperimental}) equals to
\begin{equation}
\mathrm{Var}(\widehat{\delta}_t) = \mathrm{Var}(T_{2}^{t} - T_{t+1}^{t}) = 2k\sigma ^{2} / n.
\end{equation}
For the observational approach, since each difference $T^{t}-T^{1}$ or $C^{t}-C^{1}$ is calculated in the same cohort with $n/2$ experimental units,
\begin{equation}
\begin{aligned}
\mathrm{Var}(T^{t}-T^{1}) &= 2\sigma ^{2} / n + 2\sigma ^{2} / n - 4 \rho \sigma ^{2} / n &= 4(1-\rho)\sigma ^{2} / n \\
\mathrm{Var}(C^{t}-C^{1}) &= 2\sigma ^{2} / n + 2\sigma ^{2} / n - 4 \rho \sigma ^{2} / n &= 4(1-\rho)\sigma ^{2} / n.
\end{aligned}
\end{equation}
Thus, the variance of $\widehat{\delta}_t$ in Equation~(\ref{eq-deltaobservational}) equals to
\begin{equation}
\mathrm{Var}(\widehat{\delta}_t) = \mathrm{Var}\big( (T^{t}-T^{1}) - (C^{t}-C^{1}) \big) = 8(1-\rho)\sigma ^{2} / n.
\end{equation}
Therefore, the observational approach provides more statistical power if $\rho > 1-\frac{k}{4}$ which is guaranteed if $k > 3$.

The long-term estimate of the treatment effect is the combination of the observed treatment effect during the experiment and the limit of estimated user-learning as $t \rightarrow \infty$. Sometimes, it may take a very long time for user-learning to converge. Therefore, we can use an exponential model to have an idea of how long it takes for user-learning to converge and to estimate its limit as $t \rightarrow \infty$. In Section~\ref{sec:application}, we illustrate this with a simulation study.

\subsection{Quick Detection of User-Learning} \
\label{subsec:quickDetection}
In this section we present a slightly modified version of our proposed approach to user-learning. The formulation discussed so far defines a global time window for all users in estimating user-learning. This implies that the analysis is restricted to users that are assigned to cohorts at the beginning of the experiment. To include users that are assigned to the experiment after it started, a similar formulation can be applied to user level time window: the user level time window begins when the user is exposed to the experiment and ends when the user leaves the experiment. In the calculation of $\widehat{\delta_{2}}$ in Equation~(\ref{eq-delta2observational}), we then split each user level time window into two halves. This formulation results in larger sample size and higher statistical power which leads to better inference about he population. This approach can quickly detect the significant presence of user-learning, and while is suitable for cookie-based experiments and benefits from higher statistical power, it is limited to the case with $k=3$ and may not be easily extended to the more general case with $k>3$.

We use this approach to detect user-learning in many experiments at Microsoft. This formulation is powerful in quickly testing for the presence of user-learning even in short duration experiments. In the case where user-learning is significant, we usually recommend running the experiment longer to allow for the treatment effect to stabilize\unskip~\cite{dmitriev2016pitfalls}. If the user-learning is gradually changing over time, we recommend running the experiment long enough and utilizing observational approach to construct user-learning time series that can be extrapolated to estimate the long-term treatment effect.

For illustration, we now revisit the msn.com experiment in Figure~\ref{f-5dabc0fe04f5}. Following the aforementioned formulation and the calculation of $\widehat{\delta_{2}}$ in Equation~(\ref{eq-delta2observational}), we are able to successfully detect the presence of novelty effect in total number of clicks in Figure~\ref{f-totalclicks} with a statistically significant $\widehat{\delta_2} < 0$ ($\textit{p-value} = 0.0083$).

\subsection{Discussion} \
In this section we discuss the advantages and disadvantages of both existing and proposed methodologies. The main advantage of our proposed methodology is that it provides a practically more effective way to estimate user-learning without any changes in the experimentation system. Further, as discussed above, this approach provides more statistical power for testing the significance of user-learning compared to the existing approach. The main disadvantage of our proposed methodology is that it is susceptible to treatment effect interaction with other external factors (e.g., seasonality) that are not related to user-learning. When there is a significant interaction between the treatment effect and an external factor (e.g., the treatment effect on weekend is different from the weekdays), our estimates are biased representations of user-learning. In this case, although our approach provides an unbiased estimate of the long-term treatment effect, it does not distinguish between user-learning and other forms of treatment interaction with time. Practically in controlled experiments, having a large treatment and seasonality interaction effect that significantly biases user-learning estimation is rare. Further, such effects can be avoided by running the experiment for longer period or re-running the experiment at a different time period. More complicated techniques such as time series decomposition of seasonality, cyclicality, and trend on user-learning estimates can also be used to reduce the bias in these situations. Note while we assume a linear form for $\mu_t$ in Equation~(\ref{eq-userlearningassumption}), the time series $\widehat{\delta}_t$'s can have any functional form. Thus, complex interactions of treatment effects with external factors can be modeled by more advanced methods, e.g., piecewise linear regression.

There are some known limitations for both methodologies. First, the estimates of user-learning are biased when experimental units are not durable for the period of analysis (e.g., cookie churn) and can be reset by users leading to their random movements between the cohorts\unskip~\cite{hohnhold2015focusing,dmitriev2016pitfalls}. This issue can be mitigated by restricting the analysis to experimental units that are assigned to cohorts at the beginning of the experiment\unskip~\cite{hohnhold2015focusing}. Second, the long-term estimate of the treatment effect can be biased due to violations of stable treatment unit value assumption\unskip~\cite{imbens2015causal}. For example, the behavior of one experimental unit may be influenced by another one because of social network effects or because the same user shows up as multiple experimental units in the experiment (e.g., cookies from different browsers or multiple devices). Third, the current feature change may interact with other features changes of the future product or competing products which could impact the treatment effect\unskip~\cite{czitrom1999one,wu2011experiments}. Lastly, there can be external factors that cause user behavior changes that are not captured in experiments, e.g., changes in user behavior due to COVID-19.

%\clearpage
\section{Application and Simulation} \
\label{sec:application}
In this section, we first illustrate our proposed observational approach for user-learning estimation in another real-world Microsoft experiment\footnote{The empirical examples shared in this paper are just a small selection of cases where we have observed significant user-learning.}. We then provide comparison with the existing experimental approach using a simulation study.

\subsection{Empirical Example} \
We share another example of an experiment in a Microsoft application. The feature change of this experiment impacts first launch experience after the application is updated. The treatment shows a special page informing users about the changes in the application while control shows the regular page with a notification about the update on the corner. Subsequent launches of the application show the regular page for both treatment and control. The experiment runs for about less than a month and there are about $300$K users in each control and treatment. The metric of interest is page views of the regular page where we expect a significant decrease as the result of treatment. Table~\ref{tw-microsaoftapplication} includes the results of this experiment. We utilize the simplest DID setting discussed in Section~\ref{subsec:quickDetection} to report $\widehat{\delta_{2}}$.

\begin{table}[!htbp]
\caption{{Empirical results of Microsoft application experiment} }
\label{tw-microsaoftapplication}
\def\arraystretch{1}
\ignorespaces 
\centering 
\begin{tabulary}{\linewidth}{p{\dimexpr.2\linewidth-2\tabcolsep}p{\dimexpr.25\linewidth-2\tabcolsep}p{\dimexpr.25\linewidth-2\tabcolsep}p{\dimexpr.2\linewidth-2\tabcolsep}}
\hline
Time Period & ${\widehat{\tau}}$ in $\%$ ($\textit{p-value}$) & $\widehat{\delta_{2}}$ in $\%$ ($\textit{p-value}$) & \\
\hline
First 3 days & $-5.07\%$ ($1 e-5$) & $33\%$ ($0.0027$) & $*$ \\
Week 1 &  $-3.11\%$ ($4 e-4$) & $30.30\%$ ($7 e -22$) & $*$ \\
Week 2 & $-1.54\%$ ($0.073$) & $-2.20\%$ ($0.438$) & \\
Week 3 & $-1.23\%$ ($0.163$) & $1.11\%$ ($0.620$) & \\
\hline
\end{tabulary}\par 
\end{table}

We observe a significant drop in number of page views in the first week of experiment. However, the magnitude of treatment effect is decreasing rapidly over time, and by the end of the third week, it is not statistically significant. Indeed, $\widehat{\delta_{2}}$ detects user-learning in the first week of the experiment indicating that the rate of change in treatment effect is statistically significant. By the end of the third week, neither the treatment effect nor user-learning is significant which means the feature change did not have any long-term impact on page views.

\subsection{Simulation Study} \
To compare the observational and experimental approaches, we conduct a simulation study where we use the data from the aforementioned Microsoft application experiment, and split it randomly into treatment and control cohorts. We then use a Gaussian distribution that has a mean of $\alpha e^{-\beta t}$ (with $\alpha = 1$ and $\beta = 1/3$) and standard deviation of 2 to inject a treatment effect into treatment cohorts. This simulation is consistent with the model and setup used by\unskip~\citet{hohnhold2015focusing} where they refer to the experimental approach as cookie-cookie-day. We run the simulated experiment for 14 days. Figure~\ref{f-b6d544c3e538} shows the estimated user-learning for both the observational and experimental approaches. 

\bgroup
\fixFloatSize{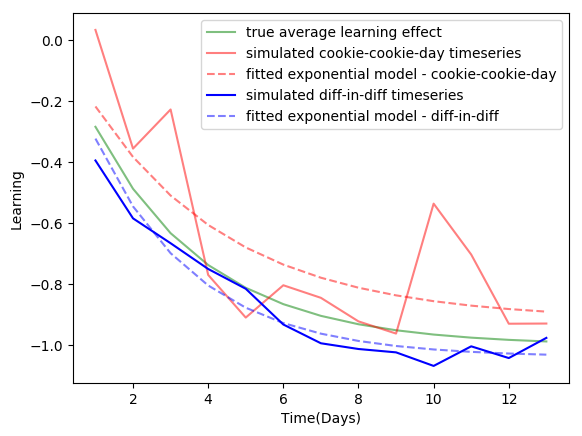}
\begin{figure}[!htbp]
\centering \makeatletter\IfFileExists{Figures/empiricalresults.png}{\includegraphics[width=.6\linewidth]{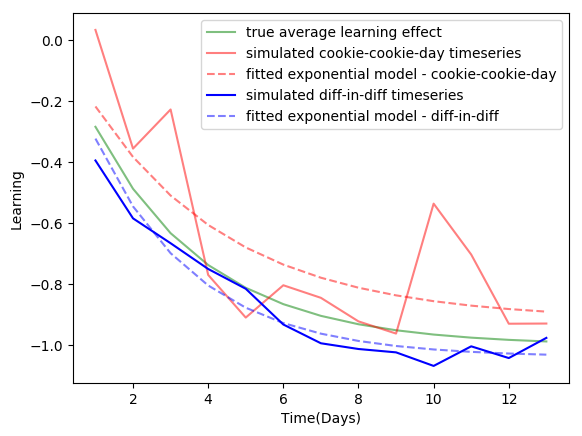}}{}
\makeatother 
\caption{{Estimated user-learning using observational and experimental approaches}}
\label{f-b6d544c3e538}
\end{figure}
\egroup

We also fit an exponential model on the time series generated by both approaches to estimate $\alpha$ and $\beta$. Further we bootstrap to estimate the standard error of these estimates. We run the simulation 1000 times with random splitting of users into different groups each time. Table~\ref{tw-simulationresult} shows the estimated values and the standard errors. As discussed in Section~\ref{sec:proposedMethodology}, the standard error is higher in the experimental approach because it requires 14 cohorts for estimation compared to 2 cohorts required in the observational approach.
 
\begin{table}[!htbp]
\caption{{Estimates of $\alpha$ and $\beta$ in the simulation study} }
\label{tw-simulationresult}
\def\arraystretch{1}
\ignorespaces 
\centering 
\begin{tabulary}{\linewidth}{p{\dimexpr.25\linewidth-2\tabcolsep}p{\dimexpr.3\linewidth-2\tabcolsep}p{\dimexpr.35\linewidth-2\tabcolsep}}
\hline Method & $\alpha $ & $\beta $\\
\hline 
True Value &
  1 &
  1/3\\
Observational  &
  1.001 (std. err.: 0.049) &
  0.339 (std. err. : 0.052)\\
Experimental &
  1.013 (std. err.: 0.105) &
  0.581 (std. err.: 2.107)\\
\hline 
\end{tabulary}\par 
\end{table}

%\clearpage
 \section{Conclusion} \
\label{sec:conclusion}
Online experiments (e.g., A/B tests) are the gold standard for evaluating impact on user experience in websites, mobile and desktop applications, services, and operating systems. At Microsoft, the experimentation system supports A/B testing across many products including Bing, Cortana, Microsoft News, Office, Skype, Windows and Xbox, running thousands of experiments per year. The metric changes observed during these experiments (typically few weeks or few months) are not always permanently stable, sometimes revealing increasing or decreasing patterns over time. There are multiple causes for a treatment effect to change over time. In this paper we focus on one particular cause, user-learning, which is primarily associated with novelty or primacy. Novelty describes the desire to use new technology that tends to diminish over time. On the contrary, primacy describes the growing engagement with technology as a result of adoption of the innovation. User-learning estimation and understanding the sustained impact of the treatment effect is critical for many reasons. First, it holds experimentation responsible for preventing overestimation or underestimation in the case of novelty or primacy. Second, it empowers organizations to make better decisions by providing them a long-term view of expected changes in the key metrics. Third, it ensures that the experiment is not causing user dissatisfaction even though the key metrics might have moved in the positive direction.

In this paper, we first formulate the problem and discuses a natural way to visually check for the presence of user-learning. We then review the existing experimental approach used in industry for user-learning estimation. This approach provides an unbiased estimate of user-learning by adding a significant operational changes to the experimentation system. It also requires a large pool of experimental units at the beginning of the experiment to be randomly divided into multiple cohorts and to be assigned to treatment in a ladder form. This approach is usually used for select few experiments where the feature team suspects user-learning a priori and is willing to have a complex experimental design setting to estimate it. It is practically more effective to estimate user-learning without any changes in the experimentation system.

We propose an observational approach, based on difference-in-differences to estimate user-learning at scale. We use this approach to detect user-learning in many experiments at Microsoft. Our formulation is powerful in quickly testing for the presence of user-learning even in short duration experiments. The main advantage of our proposed methodology is that it provides a practically more effective way to estimate user-learning by eliminating the need for the aforementioned experimental design setting. Additionally, our proposed approach provides more statistical power for testing the significance of user-learning compared to the existing approach. We further illustrate this with a simulation study. The main disadvantage of our proposed methodology is that, although it provides an unbiased estimate of the long-term treatment effect, user-learning estimation is more susceptible to other forms of treatment interaction with time (e.g., seasonality). Practically in controlled experiments, having a large treatment and seasonality interaction effect that significantly biases user-learning estimation is rare. Further, more advanced techniques such as time series decomposition of seasonality can be used to reduce the bias in user-learning estimation. 

In general, we recommend using observational approach to test for the presence of user-learning. In the case where user-learning is significant, we usually recommend running the experiment longer to allow for the treatment effect to stabilize. If the user-learning is gradually changing over time, we recommend running the experiment long enough and utilizing observational approach to construct user-learning time series that can be extrapolated to estimate the long-term treatment effect. In cases, where we suspect strong seasonality interaction with the treatment effect, and the feature team is willing to use a larger sample size with more complicated setting, experimental approach can be useful.

%\clearpage
\section*{Acknowledgements} \
\label{sec:acknowledgements}
We want to acknowledge our colleagues within Microsoft who have reviewed our work and gave valuable feedback. We also want to thank our colleagues in Microsoft Experimentation Platform team, Windows Experimentation team, and Microsoft News team for supporting our work.

\newpage
%% HERE WE DECLARE THE BIBLIOGRAPHYSTYLE TO USE AND THE BIBLIOGRAPHY DATABASE
\bibliographystyle{ECA_jasa}
\bibliography{refrences}

\end{document}